\documentclass[superscriptaddress,prl,amsmath,twocolumn,amssymb]{revtex4}

\usepackage{dcolumn}
\usepackage{amsfonts}
\usepackage{amssymb}
\usepackage{amsmath}
\usepackage{amsthm}
\usepackage{latexsym}
\usepackage{epsfig}
\usepackage{color}
\usepackage{lscape}
\usepackage{multirow}
\usepackage{braket}
\usepackage{bbm}
\usepackage{mathrsfs}
\usepackage{fancyhdr}
\usepackage{lastpage}
\usepackage{soul}
\usepackage{subfigure}
\usepackage[hidelinks=true]{hyperref}

\newcommand{\be}{\begin{equation}}
\newcommand{\ee}{\end{equation}}
\newcommand{\ba}{\begin{array}}
\newcommand{\ea}{\end{array}}
\newcommand{\bea}{\begin{eqnarray}}
\newcommand{\eea}{\end{eqnarray}}

\DeclareMathOperator{\poly}{poly}
\begin{document}

\title{Efficient Quantum Walk on a Quantum Processor}
\author{Xiaogang Qiang}
\altaffiliation{Contributed equally to this work.}
\affiliation{Centre for Quantum Photonics, H. H. Wills Physics Laboratory
and Department of Electrical and Electronic Engineering, University
of Bristol, Bristol BS8 1UB, UK.}
\author{Thomas Loke}
\altaffiliation{Contributed equally to this work.}
\affiliation{School of Physics, The University of Western Australia, WA6009, Australia.}
\author{Ashley Montanaro}
\affiliation{School of Mathematics, University of Bristol, Bristol BS8 1TW, UK.}
\author{Kanin Aungskunsiri}
\affiliation{Centre for Quantum Photonics, H. H. Wills Physics Laboratory
and Department of Electrical and Electronic Engineering, University
of Bristol, Bristol BS8 1UB, UK.}
\author{Xiaoqi Zhou}
\affiliation{Centre for Quantum Photonics, H. H. Wills Physics Laboratory
and Department of Electrical and Electronic Engineering, University
of Bristol, Bristol BS8 1UB, UK.}
\affiliation{State Key Laboratory of Optoelectronic Materials and Technologies and School of Physics and Engineering,
Sun Yat-sen University, Guangzhou 510275, China.}
\author{Jeremy L. O'Brien}
\affiliation{Centre for Quantum Photonics, H. H. Wills Physics Laboratory
and Department of Electrical and Electronic Engineering, University
of Bristol, Bristol BS8 1UB, UK.}
\author{Jingbo Wang}
\email{jingbo.wang@uwa.edu.au}
\affiliation{School of Physics, The University of Western Australia, WA6009, Australia.}
\author{Jonathan C. F. Matthews}
\email{jonathan.matthews@bristol.ac.uk}
\affiliation{Centre for Quantum Photonics, H. H. Wills Physics Laboratory
and Department of Electrical and Electronic Engineering, University
of Bristol, Bristol BS8 1UB, UK.}
\date{\today}

\begin{abstract}
\noindent The random walk formalism is used across a wide range of applications, from modelling share prices to predicting population genetics. Likewise quantum walks have shown much potential as a framework for developing new quantum algorithms. 
In this paper, we present explicit efficient quantum circuits for implementing continuous-time quantum walks on the circulant class of graphs.
These circuits allow us to sample from the output probability distributions of quantum walks on circulant graphs efficiently. We also show that solving the same sampling problem for arbitrary circulant quantum circuits is intractable for a classical computer, assuming conjectures from computational complexity theory. This is a new link between continuous-time quantum walks and computational complexity theory and it indicates a family of tasks which could ultimately demonstrate quantum supremacy over classical computers. As a proof of principle we have experimentally implemented the proposed quantum circuit on an example circulant graph using a two-qubit photonics quantum processor.
\end{abstract}

\maketitle

\noindent Quantum walks are the quantum mechanical analogue to the well-known classical random walk and they have established roles in quantum information processing~\cite{farhi1998quantum, kempe2003quantum,  childs2013universal}. In particular, they are central to quantum algorithms created to tackle database search~\cite{childs2004spatial}, graph isomorphism~\cite{douglas2008, gamble2010two, berry2011}, network analysis and navigation~\cite{Berry2010, sanchez2012quantum}, and quantum simulation~\cite{lloyd1996universal, berry2012black,schreiber20122d}, as well as modelling biological processes~\cite{engel2007evidence,rebentrost2009environment}. Meanwhile, physical properties of quantum walks have been demonstrated in a variety of systems, such as nuclear magnetic resonance~\cite{du2003experimental, ryan2005experimental}, bulk~\cite{do2005experimental} and fiber~\cite{schreiber2010photons} optics, trapped ions~\cite{xue2009quantum,schmitz2009quantum,zahringer2010realization}, trapped neutral atoms~\cite{karski2009quantum}, and photonics~\cite{perets2008realization,carolan2014experimental}. Almost all physical implementations of quantum walk so far followed an analog approach as for quantum simulation~\cite{qwbook2014, qubitImplementationQW}, whereby the apparatus is dedicated to implement specific instances of Hamiltonians without translation onto quantum logic. However, there is no existing method to implement analog quantum simulations with error correction or fault tolerance, and they do not scale efficiently in resources when simulating broad classes of large graphs.  

In this paper, we present efficient quantum circuits for implementing continuous time quantum walks (CTQWs) on circulant graphs with an eigenvalue spectrum that can be classically computed efficiently. These quantum circuits provide the time-evolution states of CTQWs on circulant graphs exponentially faster than best previously known methods~\cite{ng2004iterative}. We report a proof-of-principle experiment, where we implement CTQWs on an example circulant graph (namely the complete graph of four vertices) using a two-qubit photonics quantum processor to sample the probability distributions and perform state tomography on the output state of a CTQW. We also provide evidence from computational complexity theory that the probability distributions that are output from the circuits of this circulant form are hard to sample from using a classical computer,  implying our scheme also provides an exponential  speedup for sampling. 

\begin{figure}[t!]
\includegraphics[scale = 0.50]{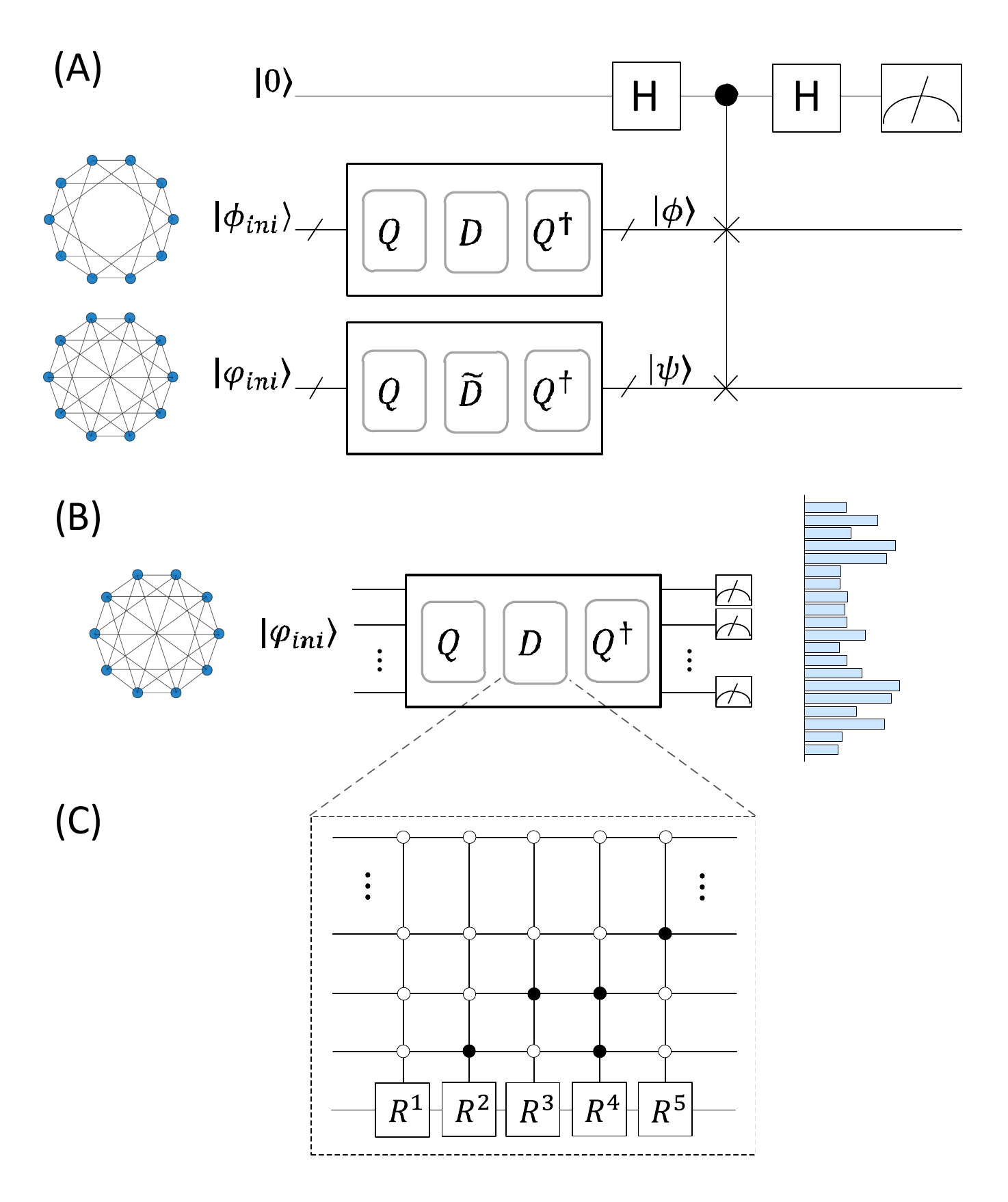}
\caption{
\footnotesize{\textbf{Applications for generating the time evolution state of circulant Hamiltonians.}
(A) The SWAP test~\cite{buhrman2001quantum} can be used to estimate the similarity of two evolution states of two similar circulant systems, or when one of the Hamiltonians is non-circulant but efficiently implementable.
In brief, an ancillary qubit is entangled with the output states $\psi$ and $\phi$ of two compared processes according to
$\frac{1}{2}\left| 0 \right\rangle \left[ {\left| \phi  \right\rangle \left| \psi  \right\rangle  + \left| \psi  \right\rangle \left| \phi  \right\rangle } \right] + \frac{1}{2}\left| 1 \right\rangle \left[ {\left| \phi  \right\rangle \left| \psi  \right\rangle  - \left| \psi  \right\rangle \left| \phi  \right\rangle } \right]$. On measuring the ancillary qubit we obtain outcome ``1'' with probability $\frac{1}{2}(1-\left|\left\langle \phi | \psi \right\rangle\right|^2)$---the probability of observing ``1'' indicates the similarity of dynamical behaviors of the two processes. 
(B) Probability distributions are sampled by measuring the evolution state in a complete basis, such as the computational basis.
(C) An example of the quantum circuit for implementing diagonal unitary operator $D=\exp( - it\Lambda )$, where the circulant Hamiltonian has $5$ non-zero eigenvalues. The open and solid circles represent the control qubits as ``if $\left| 0 \right\rangle$'' and ``if $\left| 1 \right\rangle$'' respectively. $R^i = \left[ 1,0;0,\exp({-it\lambda_i}) \right] (i=1,\cdots, 5) $, where $\lambda_{i}$ is the corresponding eigenvalue.
}}
\label{CSwapTest_Sampling}
\end{figure}

Efficient quantum circuit implementations of CTQWs have been presented for sparse and efficiently row-computable graphs~\cite{aharonov2003adiabatic,berry2007efficient}, and specific non-sparse graphs~\cite{childs2009limitations,childs2010relationship}. However, the design of quantum circuits for implementing CTQWs is in general difficult, since the time-evolution operator is time-dependent and non-local~\cite{farhi1998quantum}. 
A subset of circulant graphs have the property that their eigenvalues and eigenvectors can be classically computed efficiently~\cite{gray2006toeplitz,ng2004iterative}. This enables us to construct a scheme that efficiently outputs the quantum state $\left| {\psi \left( t \right)} \right\rangle$, which corresponds to the time evolution state of a CTQW on corresponding graphs. One can then either  perform direct measurements on $\left| {\psi \left( t \right)} \right\rangle $ or implement further quantum circuit operations to extract physically meaningful information. For example the ``SWAP test''~\cite{buhrman2001quantum} can be used to estimate the similarity of dynamical behaviors of two circulant Hamiltonians operating on two different initial states, as shown in Figure~\ref{CSwapTest_Sampling}(A). This procedure can also be adapted to study the stability of quantum dynamics of circulant molecules (for example, the DNA M\"obius strips~\cite{han2010folding}) in a perturbational environment~\cite{peres1984stability,prosen2002stability}.

On the other hand, when measuring $\left| {\psi \left( t \right)} \right\rangle $ in the computational basis we can sample the probability distribution
\begin{align}
p(x): = {\left| {\left\langle {x}
 \mathrel{\left | {\vphantom {x {\psi (t)}}}
 \right. \kern-\nulldelimiterspace}
 {{\psi (t)}} \right\rangle } \right|^2}
\end{align}
that describes the probability of observing the quantum walker at position $x \in {\left\{ {0,1} \right\}^n}$---an $n$-bit string, corresponding to the vertices of the given graph, as shown in Figure~\ref{CSwapTest_Sampling}(B).
Sampling of this form is sufficient to solve various search and characterization problems~\cite{childs2004spatial,sanchez2012quantum}, and can be used to deduce critical parameters of the quantum walk, such as mixing time~\cite{kempe2003quantum}.
It is unlikely for a classical computer to be able to efficiently sample from $p(x)$. We adapt the similar methodology of refs.~\cite{aaronson2011computational,bremner2010classical,bremner2015average} to show that if there did exist a classical sampler for a somewhat more general class of circuits,
this would have the following unlikely complexity-theoretic implication: the infinite tower of complexity classes known as the polynomial hierarchy would collapse. This evidence of hardness exists despite the classical efficiency with which properties of the CTQW, such as the eigenvalues of circulant graphs, can be computed on a classical machine. 

For an undirected graph $G$ of $N$ vertices, a quantum particle (or ``quantum walker'') placed on $G$ evolves into a superposition $\left| \psi(t)\right\rangle$ of states in the orthonormal basis $\left\{ {\left| 1 \right\rangle ,\left| 2 \right\rangle , \dots ,\left| N \right\rangle } \right\}$ that correspond to vertices of $G$. The exact evolution of the CTQW is governed by  connections between the vertices of $G$: $\left| {\psi \left( t \right)} \right\rangle = \exp( - itH)\left| {\psi \left( 0 \right)} \right\rangle$ where the Hamiltonian is given by $H = \gamma A$ for hopping rate per edge per unit time $\gamma$ and where $A$ is the $N$-by-$N$ symmetric adjacency matrix, whose entries are ${A_{jk}} = {\rm{ }}1$, if vertices $j$ and $k$ are connected by an edge in $G$, and ${A_{jk}} = {\rm{ }}0$ otherwise~\cite{farhi1998quantum}.

\begin{figure*}[t!]
\includegraphics[scale = 0.32]{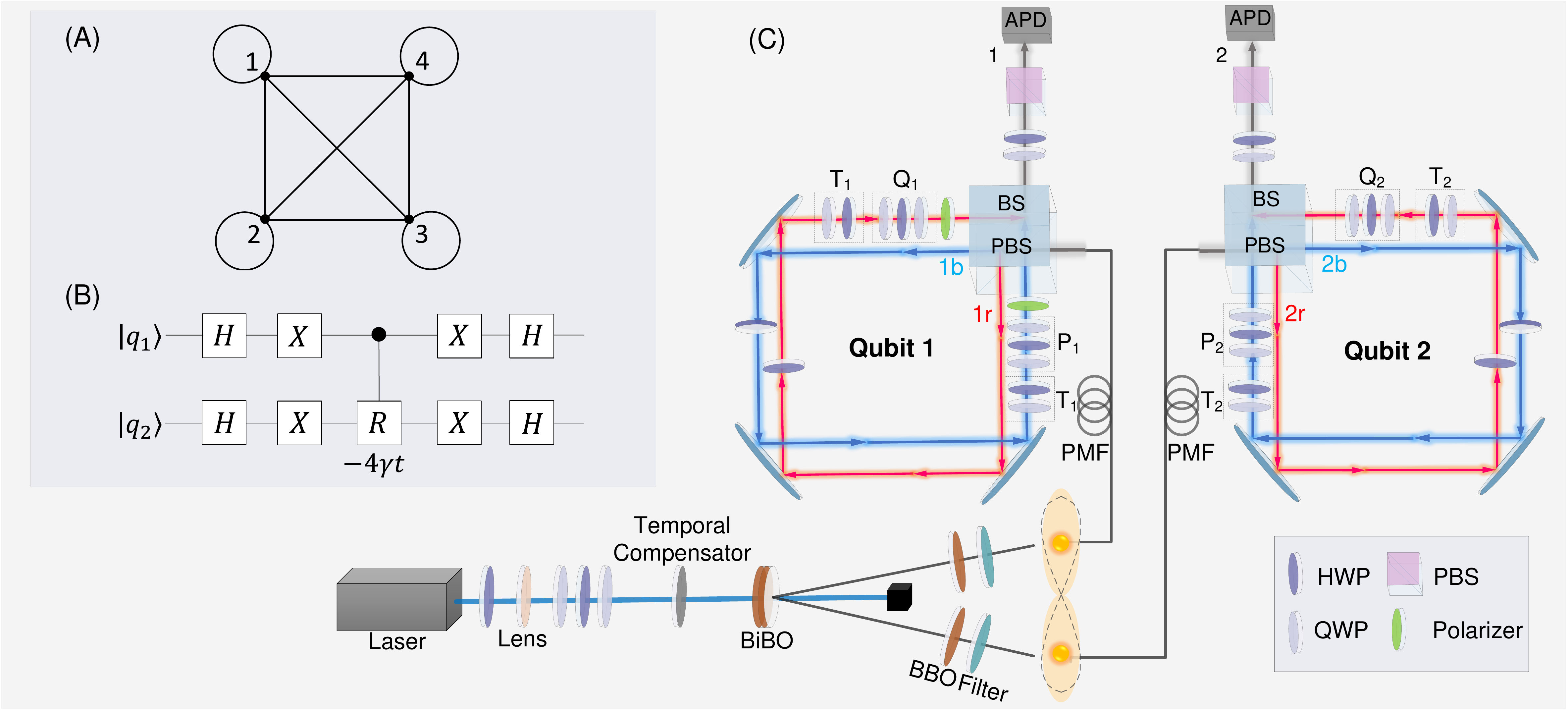}
\caption{
\footnotesize{\textbf{The schematic diagram and setup of experimental demonstration.}
(A) The $K_4$ graph.  (B) The quantum circuit for implementing CTQW on the $K_4$ graph. This can also be used to implement CTQW on the $K_4$ graph without self-loops, up to a global phase factor $\exp(i\gamma t)$. H and X represent the Hadamard and Pauli-X gate respectively. ${R} = \left[ {1,0;0,\exp(-i4\gamma t)} \right]$ is a phase gate. (C) The experimental setup for a reconfigurable two-qubit photonics quantum processor, consisting of a polarization-entangled photon source using paired type-I BiBO crystal in the sandwich configuration and displaced Sagnac interferometers. See further details in Methods.
}}
\label{ExpSetup}
\end{figure*}

Circulant graphs are defined by symmetric circulant adjacency matrices for which each row $j$ when right-rotated by one element, equals the next row $j+1$---for example complete graphs, cycle graphs and Mobius ladder graphs are all subclasses of circulant graphs.
It follows that Hamiltonians for CTQWs on any circulant graph have a symmetric circulant matrix representation, which can be diagonalized by the unitary Fourier transform~\cite{gray2006toeplitz}, i.e. $H = {Q^\dag }\Lambda Q$, where
\begin{eqnarray}
{Q_{jk}} = \frac{1}{{\sqrt N }}{\omega ^{jk}},\,\,\omega  = \exp (2\pi i/N)
\end{eqnarray}
and $\Lambda $ is a diagonal matrix containing eigenvalues of $H$, which are all real and whose order is determined by the order of the eigenvectors in $Q$. Consequently, we have $\exp( - itH ) = {Q^\dag }\exp( - it\Lambda )Q$, where the time dependence of $\exp( - itH )$ is confined to the diagonal unitary operator $D =\exp( - it\Lambda )$.

The Fourier transformation $Q$ can be implemented efficiently by the well-known QFT quantum circuit~\cite{nielsen2010quantum}. For a circulant graph that has $N=2^n$ vertices, the required QFT of $N$ dimension can be implemented with $O((\log {N})^2)=O(n^2)$ quantum gates acting on $O(n)$ qubits. To implement the inverse QFT, the same circuit is used in reverse order with phase gates of opposite sign. $D$ can be implemented using at most $N$ controlled-phase gates with phase values being a linear function of $t$, because an arbitrary phase can be applied to an arbitrary basis state, conditional on at most $n-1$ qubits.
Given a circulant graph that has $O(\poly(n))$ non-zero eigenvalues, only $O(\poly(n))$ controlled-phase gates are needed to implement $D$.
If the given circulant graph has $O(2^n)$ distinct eigenvalues, which can be characterised efficiently (such as the cycle graphs and Mobius ladder graphs), we are still able to implement the diagonal unitary operator $D$ using polynomial quantum resources. A general construction of efficient quantum circuits for $D$ was given by Childs~\cite{childs2004quantum}, and is shown in the Appendix for completeness. Thus, the quantum circuit implementations of CTQWs on circulant graphs can be constructed, which have an overall complexity of $O(\poly (n))$, and act on at most $O(n)$ qubits.
Compared with the best known classical algorithm based on fast Fourier transform, that has the computational complexity of $O(n2^n)$~\cite{ng2004iterative}, the proposed quantum circuit implementation generates the evolution state $\left| {\psi \left( t \right)} \right\rangle $ with an exponential advantage in speed.

Consider a circuit of the form $Q^\dag D Q$, where $D$ is a diagonal matrix made up of $\poly(n)$ controlled-phase gates. 
Define
\begin{align}
p_D :&=| \left\langle 0 \right|^{\otimes n} {Q^\dag }DQ\left| 0\right\rangle^{\otimes n}|^2 \nonumber \\
      & = |\left\langle +\right|^{\otimes n} D\left| +\right\rangle^{\otimes n} |^2 \nonumber \\
      & = |\left\langle 0\right|^{\otimes n} {H^{ \otimes n}}D{H^{ \otimes n}}\left| 0 \right\rangle^{\otimes n} |^2.
\end{align}
${H^{ \otimes n}}D{H^{ \otimes n}}$ represents a family of circuits having the following structure: each qubit line begins and ends with a Hadamard ($H$) gate, and, in between, every gate is diagonal in the computational basis. This class of circuits is known as instantaneous quantum polynomial time (IQP)~\cite{shepherd2009temporally,bremner2010classical}. It is known that computing $p_D$ for arbitrary diagonal unitaries $D$ made up of circuits of $\poly(n)$ gates, even if each acts on $O(1)$ qubits, is \#P-hard~\cite{bremner2015average,fujii13,goldberg14}.
This hardness result even holds for approximating $p_D$ up to any multiplicative error strictly less than $1/2$~\cite{fujii13,goldberg14}, where $\widetilde{p_D}$ is said to approximate $p_D$ up to multiplicative error $\epsilon$ if
\begin{align}
 |\widetilde{p_D} - p_D| \le \epsilon\,p_D.
\end{align}

Towards a contradiction, assume that there exists a polynomial-time randomized classical algorithm which samples from $p$. Then a classic result of Stockmeyer~\cite{stockmeyer1985on} states that there is an algorithm in the complexity class FBPP$^{\text{NP}}$ which can approximate any desired probability $p(x)$ to within multiplicative error $O(1/\poly(n))$. This complexity class FBPP$^{\text{NP}}$---described as polynomial-time randomized classical computation equipped with an oracle to solve arbitrary NP problems---sits within the infinite tower of complexity classes known as the polynomial(-time) hierarchy~\cite{papadimitriou1994computational}. Combining with the above hardness result of approximating $p_D$, we find that the assumption implies that 
an FBPP$^\text{NP}$ algorithm solves a \#P-hard problem, so P$^\text{\#P}$ would be contained within FBPP$^\text{NP}$, and therefore the polynomial hierarchy would collapse to its third level. This consequence is considered very unlikely in computational complexity theory~\cite{papadimitriou1994computational}.

We therefore conclude that a polynomial-time randomized classical sampler from the distribution $p$ is unlikely to exist. Further, this even holds for classical algorithms which sample from any distribution $\widetilde{p}$ which approximates $p$ up to multiplicative error strictly less than $1/2$ in each probability $p(x)$. It is worth noting that if the output distribution results from measurements on only $O(\poly \log n)$ qubits~\cite{van2011simulating}, or obeys the sparsity promise that only a $\poly(n)$-sized, and a priori unknown, subset of the measurement probabilities are nonzero~\cite{schwarz2013simulating}, it could be classically efficiently sampled. The proof of hardness here does not hold for the approximate sampling from $p$ up to small {\em additive} error. It is an interesting open question whether similar techniques to~\cite{aaronson2011computational,bremner2015average} can be used to prove hardness of the approximate case, perhaps conditioned on other conjectures in complexity theory.

\begin{figure*}[hbt]
\includegraphics[scale = 0.5]{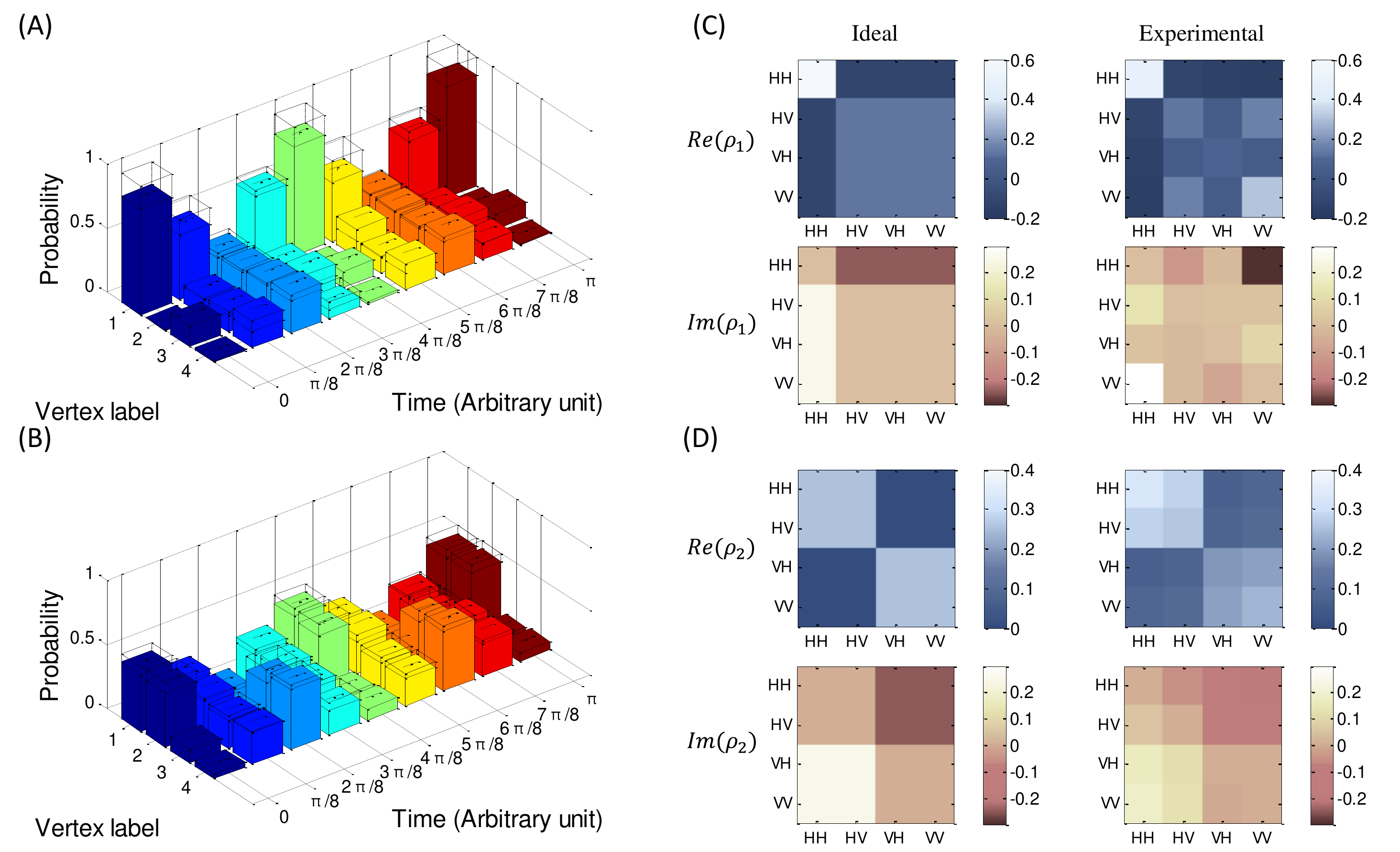}
\caption{
\footnotesize{\textbf{Experimental results for simulating CTQWs on $K_4$.}
(A)-(B) The experimental sampled probability distributions with ideal theoretical distributions overlaid, for CTQWs on $K_4$ graph with initial states ${\left| {{\varphi _{ini}}} \right\rangle _1} = \left[ {1,0,0,0} \right]'$ and ${\left| {{\varphi _{ini}}} \right\rangle _2} = \frac{1}{{\sqrt 2 }}\left[ {1,1,0,0} \right]'$. The standard deviation of each individual probability is also plotted, which is calculated by propagating error assuming Poissonian statistics. (C)-(D) The ideal theoretical and experimentally reconstructed density matrices for the states ${\left| {{\varphi _{out}}} \right\rangle _1} = \left[ {0.75 + 0.25i, - 0.25 + 0.25i, - 0.25 + 0.25i, - 0.25 + 0.25i} \right]'$ (corresponding to $\rho _1$) and ${\left| {{\varphi _{out}}} \right\rangle _2} = \left[ {0.3536 + 0.3536i,0.3536 + 0.3536i, - 0.3536 + 0.3536i, - 0.3536 + 0.3536i} \right]'$ (corresponding to $\rho _2$). Both of the real and imaginary parts of the density matrices are obtained through the maximum likelihood estimation technique, and is shown as $\mbox{Re}(\rho)$ and $\mbox{Im}(\rho)$ respectively. Further results are shown in Appendix.
}}
\label{ExpData}
\end{figure*}

As an experimental demonstration, we used a photonic quantum logic to simulate CTQWs on the $K_4$ graph---a complete graph with self loops on four vertices (Figure~\ref{ExpSetup}(A)).
Complete graphs are a special kind of circulant graph, with an adjacency matrix $A$ where $A_{jk}=1$ for all $j, k$. The Hamiltonian of a complete graph on $N$ vertices has only 2 distinct eigenvalues, 0 and $N\gamma$. Therefore, the diagonal matrix of eigenvalues of $K_4$ is $\Lambda  = \mbox{diag}(\{ 4\gamma,0,0,0\} )$. Following the aforementioned discussion, we can readily construct the quantum circuit for implementing CTQWs on $K_4$ graph based on diagonalization using the QFT matrix. However, the choice of using the QFT matrix as the eigenbasis of Hamiltonian is not strictly necessary -- any equivalent eigenbasis can be selected. Through the diagonalization using Hadamard eigenbasis, an alternative efficient quantum circuit for implementing CTQWs on $K_4$ graph is shown in Figure~\ref{ExpSetup}(B), which can be easily extended to the complete graph on $N$ vertices.

We built a configurable two-qubit photonics quantum processor (Figure~\ref{ExpSetup}(C)), adapting the entanglement-based technique presented in~\cite{zhou2011adding}, and implemented CTQWs on $K_4$ graph with various evolving times and initial states. Specifically, we prepared two different initial states
${\left| {{\varphi _{ini}}} \right\rangle _1} = \left[ {1,0,0,0} \right]'$ and ${\left| {{\varphi _{ini}}} \right\rangle _2} = \frac{1}{{\sqrt 2 }}\left[ {1,1,0,0} \right]'$, which represent the quantum walker starting from vertex 1, and the superposition of vertices 1 and 2 respectively. We chose the evolution time following the list
$\left\{ {0,\frac{1}{8}\pi ,\frac{2}{8}\pi ,\frac{3}{8}\pi ,\frac{4}{8}\pi ,\frac{5}{8}\pi ,\frac{6}{8}\pi ,\frac{7}{8}\pi ,\pi } \right\}$,
which covers the whole periodical characteristics of CTQWs on $K_4$ graph. For each evolution, we sampled the corresponding probability distribution with fixed integration time, shown in Figure~\ref{ExpData}(A) and (B).
To measure how close the experimental and ideal probability distributions are, we calculated the average fidelities defined as
$
{F_{average}} = \frac{1}{9}\sum\nolimits_{n = 1}^9 {\sum\nolimits_{i = 1}^4 {\sqrt {{P_{ideal,n}}(i){P_{\exp ,n}}(i)} } }
$.
The achieved average fidelities for the samplings with two distinct initial states are 96.68$\pm$0.27\% and 95.82$\pm$0.25\% respectively.
Through the proposed circuit implementation, we are also able to examine the evolution states using quantum state tomography, which is generally difficult for the analog simulations.
For two specific evolution states
${\left| {{\varphi _{out}}} \right\rangle _1} = \exp({ - iH\frac{7}{8}\pi }){\left| {{\varphi _{ini}}} \right\rangle _1}$ and
${\left| {{\varphi _{out}}} \right\rangle _2} = \exp({ - iH\frac{7}{8}\pi }){\left| {{\varphi _{ini}}} \right\rangle _2}$,
we performed quantum state tomography and reconstructed the density matrices using the maximum likelihood estimation technique. The two reconstructed density matrices achieve fidelities of 85.81$\pm$1.08\% and 88.44$\pm$0.97\% respectively, shown in Figure~\ref{ExpData}(C) and (D).

In this paper, we have described how CTQWs on circulant graphs can be efficiently implemented on a quantum computer, if the eigenvalues of the graphs can be characterised efficiently classically. 
In fact, we can construct an efficient quantum circuit to implement CTQWs on any graph whose adjacency matrix is efficiently diagonalisable, in other words, as long as the matrix of column eigenvectors $Q$ and the diagonal matrix of the eigenvalue exponentials $D$ can be implemented efficiently.
We have shown that the problem of sampling from the output probability distributions of quantum circuits of the form $Q^\dag DQ$ is hard for classical computers, based on a highly plausible conjecture that the polynomial hierarchy does not collapse. 
This observation is particularly interesting from both perspectives of CTQW and computational complexity theory, as it provides new insights into the CTQW framework and also helps to classify and identify new problems in computational complexity theory.
For the CTQWs on the circulant graphs of $\mbox{poly}(n)$ non-zero eigenvalues, the proposed quantum circuit implementations do not need a fully universal quantum computer, and thus can be viewed as an intermediate model of quantum computation. Although the hardness of the approximate case of the sampling problem is unknown, the evidence we provided for the exact case indicates a promising candidate for experimentally establishing quantum supremacy over classical computers, and further evidence against the extended Church-Turing thesis. Compared with other intermediate models such as the one clean qubit model~\cite{knill1998power}, IQP and boson sampling~\cite{aaronson2011computational}, the quantum circuit implementation of CTQWs is also more appealing due to available methods in fault tolerance and error correction, which are difficult to implement for other models~\cite{rohde2012error}. This may also lead onto other practical applications through the use of CTQWs for quantum algorithm design.

\small{
\noindent \textbf{Experimental Setup}
A diagonally polarized, 120 mW, continuous-wave laser beam with central wavelength of 404 nm is focused at the centre of paired type-I BiBO crystals with their optical axes orthogonally aligned to each other, to create the polarization entangled photon-pairs~\cite{rangarajan2009optimizing}. Through the spontaneous parametric down-conversion process, the photon pairs are generated in the state of $\frac{1}{{\sqrt 2 }} \left( \left| {H_1H_2} \right\rangle  + \left| {V_1V_2} \right\rangle \right)$, where $H$ and $V$ represent horizontal and vertical polarization respectively. The photons pass through the polarization beam-splitter (PBS) part of the dual PBS/beam-splitter (BS) cubes on both arms to generate two-photon four-mode state of the form $\frac{1}{{\sqrt 2 }} \left( \left| {H_{1b}H_{2b}} \right\rangle  + \left| {V_{1r}V_{2r}} \right\rangle \right)$ ($r$ and $b$ label red and blue paths shown in Figure~\ref{ExpSetup}(C)). Rotations $T_1$ and $T_2$ on each path, consisting of half wave-plate (HWP) and quarter wave-plate (QWP), convert the state into $\frac{1}{{\sqrt 2 }} \left( \left| {\phi_{1b}\phi_{2b}} \right\rangle  + \left| {\phi_{1r}\phi_{2r}} \right\rangle \right)$, where $\left| {\phi _ 1} \right\rangle$ and $\left| {\phi _2} \right\rangle$ can be arbitrary single-qubit states. The four spatial modes $1b$, $2b$, $1r$ and $2r$ pass through four single-qubit quantum gates $P_1$, $P_2$, $Q_1$ and $Q_2$ respectively, where each of the four gates is implemented through three wave plates: QWP, HWP and QWP.
The spatial modes $1b$ and $1r$ ($2b$ and $2r$) are then mixed on the BS part of the cube.
By post-selecting the case where the two photons exit at ports 1 and 2, we obtain the state $\left( P_1 \otimes P_2 + Q_1 \otimes Q_2 \right)\left| {\phi_{1}\phi_{2}} \right\rangle $. In this way, we implement a two-qubit quantum operation of the form $P_1 \otimes P_2 + Q_1 \otimes Q_2$ on the initialized state $\left| {\phi_{1}\phi_{2}} \right\rangle$.

As shown in Figure~\ref{ExpSetup}(B), the quantum circuit for implementing CTQW on the $K_4$ graph consists of Hadamard gates (H), Pauli-X gates (X) and controlled-phase gate (CP). CP is implemented by configuring
${P_1} = \left|  H  \right\rangle \left\langle  H  \right|$,
${P_2} = I$,
${Q_1} = \left|  V  \right\rangle \left\langle  V  \right|$,
${Q_2} = R ( = \left[ {1,0;0,{e^{-i4\gamma t }}} \right])$, where $P_1$ and $Q_1$ are implemented by polarizers. Together with combining the operation $\left( {H \cdot X} \right) \otimes \left( {H \cdot X} \right)$ before CP with state preparation and the operation $\left( {X \cdot H} \right) \otimes \left( {X \cdot H} \right)$ after CP with measurement setting, we implement the whole quantum circuit on the experimental setup. The evolution time of CTQW is controlled by the phase value of $R$, which is determined by setting the three wave plates of $Q_2$ in Figure~\ref{ExpSetup}(C) to QWP($\frac{\pi }{4}$), HWP($\omega $), QWP($\frac{\pi }{4}$), where the angle $\omega $ of HWP equals to the phase $\theta$ of $R$. 
The evolution time $t$ is then given by ${{t =  - \omega } \mathord{\left/
 {\vphantom {{t =  - \omega } {\left( {4\gamma } \right)}}} \right.
 \kern-\nulldelimiterspace} {\left( {4\gamma } \right)}}.$
}
\acknowledgements


\clearpage


\pagebreak
\newpage
\begin{center}
\section*{\large Appendix}
\end{center}
\setcounter{equation}{0}
\setcounter{figure}{0}
\setcounter{table}{0}
\setcounter{page}{1}
\makeatletter
\renewcommand{\theequation}{A\arabic{equation}}
\renewcommand{\thefigure}{A\arabic{figure}}
\renewcommand{\bibnumfmt}[1]{[#1]}
\renewcommand{\citenumfont}[1]{#1}
\normalsize

\section{\textbf {Further Details on Circulant Graphs and Other Examples}}
A circulant graph of $N$ vertices is fully described by an $N$-by-$N$ symmetric circulant adjacency matrix $C$ defined as follows.
\begin{align}
C = \left[ {\begin{array}{*{20}{c}}
{{c_0}}&{{c_1}}&{{c_2}}& \ldots &{{c_{N - 1}}}\\
{{c_{N - 1}}}&{{c_0}}&{{c_1}}& \ldots &{{c_{N - 2}}}\\
{{c_{N - 2}}}&{{c_{N - 1}}}&{{c_0}}& \ldots &{{c_{N - 3}}}\\
 \vdots & \vdots & \vdots & \ddots & \vdots \\
{{c_1}}&{{c_2}}&{{c_3}}& \ldots &{{c_0}}
\end{array}} \right]
\end{align}
where $c_j = c_{N-j}, j = 1, 2, \cdots, N-1$.
Obviously, every circulant matrix can be generated given any row of the matrix -- conventionally we use the first row of the matrix, denoted as $r_C$. It is clear that $C$ has at most $N$ distinct eigenvalues which are given by ${\lambda _m} = \sum\nolimits_{k = 0}^{N - 1} {{c_k}{\omega ^{ - mk}}} $, where $\omega  = \exp({2\pi i/N})$ and $m = 0,1, \ldots ,N - 1$~\cite{ng2004iterative_appendix}. If $C$ is singular, some of the eigenvalues of $C$ are zeros.
The complete graph and complete bipartite graph are straightforward examples of circulant graphs with few distinct eigenvalues.

There are also some other interesting examples of circulant graph such as self-complementary circulant graphs and Paley graphs with prime order~\cite{zhou2011self, rajasingh2013spanning}. Both of these two families of graphs are also strongly regular graphs which have only three distinct eigenvalues. For example, the Paley graph on 13 vertices has three distinct eigenvalues: 6 (with multiplicity 1) and $\frac{1}{2}\left( { - 1 \pm \sqrt {13} } \right)$ (both with multiplicity 6), and thus the diagonal unitary $\exp(-it\Lambda)$ can be implemented efficiently.
We note here it is required to implement QFT (and its inverse) for the dimension of 13, which does not have the form of $N=2^n$. The QFT on general dimensions can be implemented by means of amplitude amplification with extra qubit registers to perform the computation~\cite{mosca2004exact}. Alternatively, approximate versions of the QFT on general dimensions have also been developed~\cite{hales2000improved}.

\section{ \textbf {Implementation of the Diagonal Unitary Operator}}
\begin{figure}[t!]
\includegraphics[scale = 0.5]{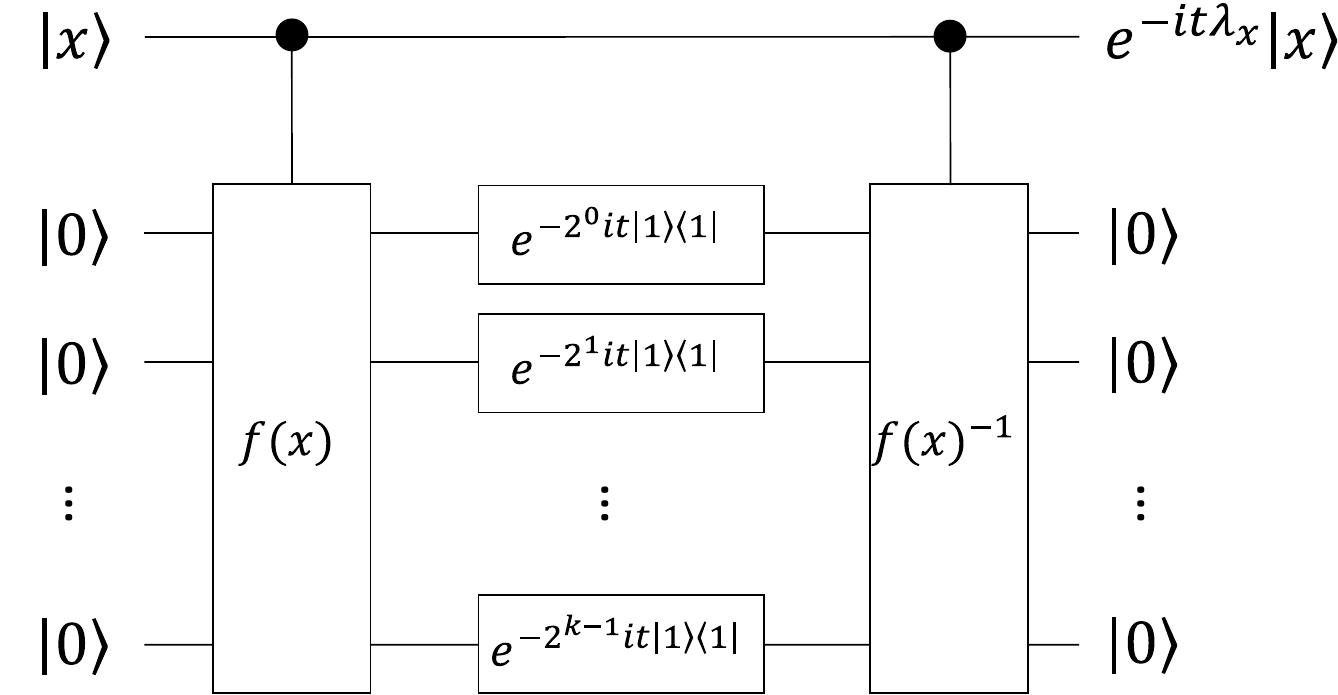}
\caption{
\footnotesize{\textbf{}
The quantum circuit for implementing the diagonal unitary operator $\exp(-it\Lambda)$ of the given circulant Hamiltonian. The eigenvalues of given circulant can be calculated by the function $f(x)$ efficiently classically.
}}
\label{DiagonalUnitaryCircuit}
\end{figure}

We say that the eigenvalues of a circulant graph can be characterised efficiently, if they can be calculated efficiently classically.
In other words, the eigenvalue matrix $\Lambda$ of the given circulant Hamiltonian can be efficiently computed, and thus the diagonal unitary operator $\exp(-it\Lambda)$ can be efficiently implemented~\cite{childs2004quantum}.
Specifically, there exists a quantum circuit shown in Figure~\ref{DiagonalUnitaryCircuit}, which transforms a computational basis state $\left| x \right\rangle$, together with a $k$-qubit ancilla $\left| 0 \right\rangle$ for $k=\poly(n)$, as
\begin{align}
\left| x \right\rangle \left| 0 \right\rangle
&\to \left| x \right\rangle \left| {{\lambda _x}} \right\rangle \nonumber \\
&\to {e^{ - it {\lambda _x}}}\left| x \right\rangle \left| {{\lambda _x}} \right\rangle  \nonumber \\
&\to {e^{ - it {\lambda _x}}}\left| x \right\rangle \left| 0 \right\rangle
= {e^{ - it\Lambda}}\left| x \right\rangle \left| 0 \right\rangle
\end{align}
where $x = 0,1, \ldots ,N - 1$. Note that here we assume $\lambda_x$ can be expressed exactly as a rational number with $k$ bits of precision. If this is not the case, truncating $\lambda_x$ to $k$ bits of precision will introduce an error which can be made arbitrarily small by taking large enough $k=\poly(n)$.
The function $f(x)$ returns $\lambda _x$ for any given $x$. $\lambda _x$ is always a real number since the adjacency matrix is symmetric.

For example, for the case of the cycle graph of $N=2^n$ vertices, there are essentially $N/2$ distinct eigenvalues simply given by ${\lambda _x} = 2\cos \left( {2\pi x/N} \right)$, where $x = 0,1, \ldots ,N - 1$. And then $f(x)$ will be the cosine function that can be computed with a number of operations polynomial in $n$, using a reversible equivalent of classical algorithms to compute trigonometric functions, e.g. the Taylor approximation.
In general, given a sparse circulant graph which has only $\poly(n)$ $1$s in the first row $r_C$ of its adjacency matrix, an efficient function $f(x)$ can be given as
\begin{align}
f(x) = \sum\limits_{y \in S} {{e^{2i\pi xy/N}}}
\end{align}
where $S$ is the set of positions for which the first row in nonzero. $f(x)$ is a sum of $\left| S \right| = \poly(n)$ numbers, taking $O(\poly(n))$ time to compute. For a non-sparse circulant graph, its eigenvalues are still possible to be calculated efficiently classically. Some straightforward examples are complete graph, complete bipartite graph $K_{N,N}$ and cocktail party graph.
Therefore, together with the quantum circuits of QFT and the inverse of QFT, we construct an efficient quantum circuit for implementing CTQW on the circulant graph whose eigenvalues can be computed efficiently classically.

\section{\textbf {Complexity Analysis of ``SWAP test''}}
Unlike the sampling problem we discussed in the main text, the scenario of ``SWAP test'', where we compare two unitary processes $Q^\dag D Q$ and $Q^\dag \widetilde{D} Q$, could sometimes be easier for a classical computer.
Imagine we start each process in the state $\left| {{\phi _{ini}}} \right\rangle = \left| {{\varphi _{ini}}} \right\rangle = \left| 0 \right\rangle^{\otimes n}$. Then the overlap $O$ between the resulting output states approximated by the SWAP test satisfies
\begin{align}
O &= |\left\langle 0\right|^{\otimes n} (Q^\dag D Q) (Q^\dag \widetilde{D} Q) \left| 0 \right\rangle^{\otimes n} |^2 \nonumber\\
&=  |\left\langle +\right|^{\otimes n} D \widetilde{D}  \left| + \right\rangle^{\otimes n} |^2
= \big| \frac{1}{2^n} \sum_{x \in \{0,1\}^n} D_{xx} \widetilde{D}_{xx} \big|^2,
\end{align}
where $D_{xx}$ is the value at position $x$ on the diagonal of $D$. $O$ can be approximated by a classical algorithm up to $O(1/\poly(n))$ additive error. The algorithm simply takes the average of $\poly(n)$ values of the product $D_{xx} \widetilde{D}_{xx}$ for uniformly random $x$. For each $x$, this value can be computed exactly in polynomial time.

This highlights that the complexity of comparing $Q^\dag D Q \left| {{\phi _{ini}}} \right\rangle$ and $Q^\dag \widetilde{D} Q \left| {{\varphi _{ini}}} \right\rangle$ depends on the choice of input states $\left| {{\phi _{ini}}} \right\rangle$ and $\left| {{\varphi _{ini}}} \right\rangle$. In full generality, one could allow these to be arbitrary states produced by a polynomial-time quantum computation; the state comparison problem would then be BQP-complete, but for rather trivial reasons. We expect that the problem would remain classically hard for choices of initial states relevant, for example, to quantum-chemistry applications. On the other hand, the SWAP test can still be used as in Scenario (B) in the main text to compare the evolution of two Hamiltonians, one of which is not circulant but is efficiently implementable. In this case, the comparison problem is also BQP-complete, and hence expected to be hard for a classical computer.

\section{ \textbf {Further Experimental Results}}
We present the ideal and experimentally sampled probability distributions of CTQW with initial states
${\left| {{\varphi _{ini}}} \right\rangle _1} = \left[ {1,0,0,0} \right]'$,
${\left| {{\varphi _{ini}}} \right\rangle _2} = \frac{1}{{\sqrt 2 }}\left[ {1,1,0,0} \right]'$
(mentioned in the main text),
${\left| {{\varphi _{ini}}} \right\rangle _3} = \frac{1}{{\sqrt 2 }}\left[ {1, - 1,0,0} \right]'$,
${\left| {{\varphi _{ini}}} \right\rangle _4} = \frac{1}{{\sqrt 2 }}\left[ {1, - i,0,0} \right]'$,
${\left| {{\varphi _{ini}}} \right\rangle _5} = \frac{1}{ 2 }\left[ {1, i, 1, i} \right]'$,
${\left| {{\varphi _{ini}}} \right\rangle _6} = \frac{1}{ 2 }\left[ {1, i, i, -1} \right]'$,
in Table I. The achieved average fidelities between ideal and experimental probability distributions are 96.68$\pm$0.27\%, 95.82$\pm$0.25\%, 92.61$\pm$0.21\%, 96.36$\pm$0.16\%, 98.76$\pm$0.17\% and 97.27$\pm$0.24\% respectively.
In the main text, we reconstructed the density matrices for the two quantum states ${\left| {{\varphi _{out}}} \right\rangle _1}$ and ${\left| {{\varphi _{out}}} \right\rangle _2}$, through performing quantum state tomography. Here we also present the reconstructed density matrices for another two evolution states
${\left| {{\varphi _{out}}} \right\rangle _3} = \exp({ - iH\frac{3}{4}\pi }){\left| {{\varphi _{ini}}} \right\rangle _1}$ and
${\left| {{\varphi _{out}}} \right\rangle _4} = \exp({ - iH\frac{3}{4}\pi }){\left| {{\varphi _{ini}}} \right\rangle _2}$, with the achieved fidelities of 88.63$\pm$1.24\% and 91.53$\pm$0.53\% respectively.
See in Figure~\ref{densityM12}.
The four reconstructed density matrices $\rho _1$, $\rho _2$, $\rho _3$ and $\rho _4$ for quantum states ${\left| {{\varphi _{out}}} \right\rangle _1}$, ${\left| {{\varphi _{out}}} \right\rangle _2}$, ${\left| {{\varphi _{out}}} \right\rangle _3}$ and ${\left| {{\varphi _{out}}} \right\rangle _4}$  are shown as follows.

\begin{widetext}

\begin{figure*}[t!]
\includegraphics[scale = 0.5]{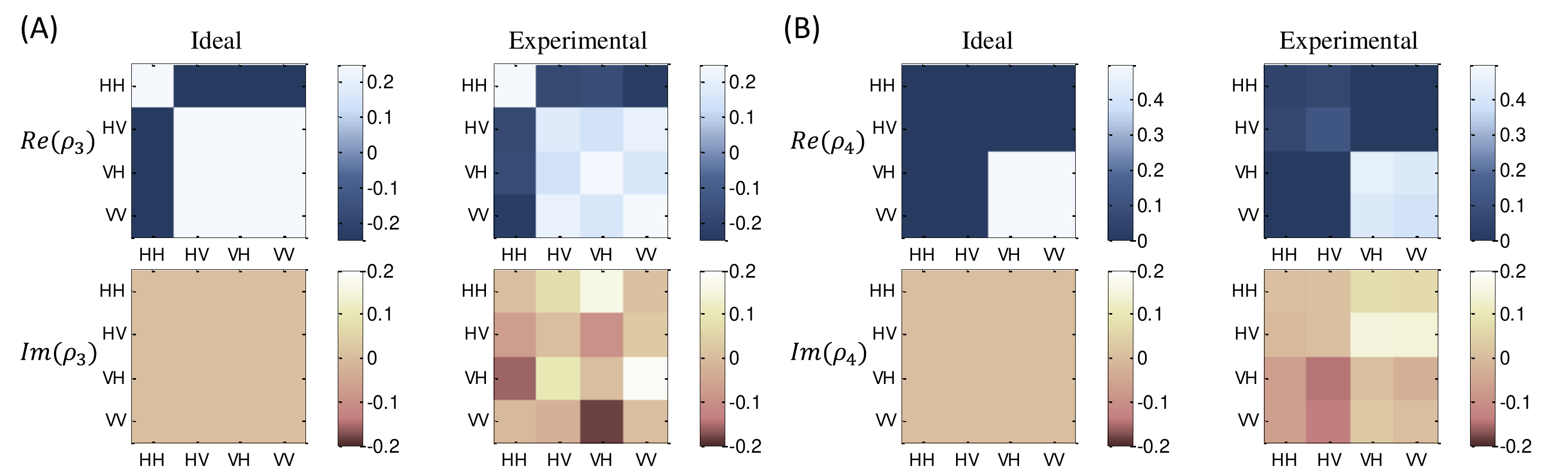}
\caption{
\footnotesize{\textbf{}
(A) The ideal theoretical and experimentally reconstructed density matrices for the states ${\left| {{\varphi _{out}}} \right\rangle _3} = \frac{1}{ 2 }\left[ {1, -1, -1, -1} \right]'$ (corresponding to $\rho _3$) and (B) ${\left| {{\varphi _{out}}} \right\rangle _4} = \frac{1}{\sqrt 2 }\left[ {0, 0, -1, -1} \right]'$ (corresponding to $\rho _4$). Both of the real and imaginary parts of the density matrices are obtained through the maximum likelihood estimation technique, and shown as $\mbox{Re}(\rho)$ and $\mbox{Im}(\rho)$ respectively. The achieved fidelities are 88.63$\pm$1.24\% and 91.53$\pm$0.43\% respectively.
}}
\label{densityM12}
\end{figure*}

\[{\rho _1} = \left[ {\begin{array}{*{20}{c}}
{0.4763}&{ - 0.1175 - 0.1281i}&{ - 0.1410 - 0.0112i}&{ - 0.1507 - 0.3104i}\\
{ - 0.1175 + 0.1281i}&{0.1354}&{0.0257 + 0.0115i}&{0.1620 + 0.0133i}\\
{ - 0.1410 + 0.0112i}&{0.0257 - 0.0115i}&{0.0841}&{0.0289 + 0.0797i}\\
{ - 0.1507 + 0.3104i}&{0.1620 - 0.0133i}&{0.0289 - 0.0797i}&{0.3041}
\end{array}} \right]\]

\[{\rho _2} = \left[ {\begin{array}{*{20}{c}}
{0.3207}&{0.2801 - 0.0479i}&{0.0665 - 0.1666i}&{0.0851 - 0.1810i}\\
{0.2801 + 0.0479i}&{0.2575}&{0.0812 - 0.1135i}&{0.0989 - 0.1245i}\\
{0.0665 + 0.1666i}&{0.0812 + 0.1135i}&{0.1899}&{0.2044 + 0.0088i}\\
{0.0851 + 0.1840i}&{0.0989 + 0.1245i}&{0.2044 - 0.0088i}&{0.2319}
\end{array}} \right]\]

\[{\rho _3} = \left[ {\begin{array}{*{20}{c}}
{0.2779}&{ - 0.1804 + 0.0749i}&{ - 0.1711 + 0.1613i}&{ - 0.2397 + 0.0091i}\\
{ - 0.1804 - 0.0749i}&{0.1778}&{0.1376 - 0.0938i}&{0.2083 + 0.0311i}\\
{ - 0.1711 - 0.1613i}&{0.1376 + 0.0938i}&{0.2411}&{0.1507 + 0.1825i}\\
{ - 0.2397 - 0.0091i}&{0.2083 - 0.0311i}&{0.1507 - 0.1825i}&{0.3031}
\end{array}} \right]\]

\[{\rho _4} = \left[ {\begin{array}{*{20}{c}}
{0.0418}&{0.0605 + 0.0067i}&{ - 0.0353 + 0.0698i}&{ - 0.0287 + 0.0671i}\\
{0.0605 - 0.0067i}&{0.1196}&{ - 0.0260 + 0.1465i}&{ - 0.0155 + 0.1381i}\\
{ - 0.0353 - 0.0698i}&{ - 0.0260 - 0.1465i}&{0.4481}&{0.4175 - 0.0263i}\\
{ - 0.0287 - 0.0671i}&{ - 0.0155 - 0.1381i}&{0.4175 + 0.0263i}&{0.3904}
\end{array}} \right]\]

\end{widetext}

\newpage

\begin{table*}[t!]
\centering
\setlength{\tabcolsep}{5pt}
\begin{tabular}{lrrrrrrrrr}
\hline
Node/T     & 0	&$\frac{1}{8}\pi $	&$\frac{2}{8}\pi $	&$\frac{3}{8}\pi $	&$\frac{4}{8}\pi $	&$\frac{5}{8}\pi $	&$\frac{6}{8}\pi $	&$\frac{7}{8}\pi $	&$\pi $\\
\hline
$P_{1,ideal}$	&	1	&	0.625	&	0.25	&	0.625	&	1	&	0.625	&	0.25	&	0.625	&	1\\
$P_{2,ideal}$	&	0	&	0.125	&	0.25	&	0.125	&	0	&	0.125	&	0.25	&	0.125	&	0\\
$P_{3,ideal}$	&	0	&	0.125	&	0.25	&	0.125	&	0	&	0.125	&	0.25	&	0.125	&	0\\	
$P_{4,ideal}$	&	0	&	0.125	&	0.25	&	0.125	&	0	&	0.125	&	0.25	&	0.125	&	0\\
$P_{1,exp}$	&	0.8225	&	0.5014	&	0.2139	&	0.5759	&	0.8482	&	0.4583	&	0.2414	&	0.549	&	0.858\\
$P_{2,exp}$	&	0.003	&	0.1388	&	0.2254	&	0.1455	&	0.0078	&	0.2167	&	0.2375	&	0.1324	&	0.0114\\
$P_{3,exp}$	&	0.1598	&	0.153	&	0.2659	&	0.2105	&	0.1284	&	0.1333	&	0.2299	&	0.1912	&	0.1193\\
$P_{4,exp}$	&	0.0148	&	0.2068	&	0.2948	&	0.0681	&	0.0156	&	0.1917	&	0.2912	&	0.1275	&	0.0114\\
\hline
\end{tabular}

\begin{tabular}{lrrrrrrrrr}
\hline
Node/T     & 0	&$\frac{1}{8}\pi $	&$\frac{2}{8}\pi $	&$\frac{3}{8}\pi $	&$\frac{4}{8}\pi $	&$\frac{5}{8}\pi $	&$\frac{6}{8}\pi $	&$\frac{7}{8}\pi $	&$\pi $\\
\hline
$P_{1,ideal}$	&	0.5	&	0.25	&	0	&	0.25	&	0.5	&	0.25	&	0	&	0.25	&	0.5\\
$P_{2,ideal}$	&	0.5	&	0.25	&	0	&	0.25	&	0.5	&	0.25	&	0	&	0.25	&	0.5\\
$P_{3,ideal}$	&	0	&	0.25	&	0.5	&	0.25	&	0	&	0.25	&	0.5	&	0.25	&	0\\
$P_{4,ideal}$	&	0	&	0.25	&	0.5	&	0.25	&	0	&	0.25	&	0.5	&	0.25	&	0\\
$P_{1,exp}$	&	0.4386	&	0.2796	&	0.0682	&	0.2927	&	0.4375	&	0.2823	&	0.0669	&	0.2338	&	0.434\\
$P_{2,exp}$	&	0.4189	&	0.2607	&	0.0746	&	0.2717	&	0.4103	&	0.3008	&	0.0605	&	0.2	&	0.4415\\
$P_{3,exp}$	&	0.1031	&	0.2156	&	0.3945	&	0.2482	&	0.0679	&	0.2058	&	0.4108	&	0.2923	&	0.0566\\
$P_{4,exp}$	&	0.0395	&	0.2441	&	0.4627	&	0.1874	&	0.0842	&	0.2111	&	0.4618	&	0.2738	&	0.0679\\
\hline
\end{tabular}

\begin{tabular}{lrrrrrrrrr}
\hline
Node/T     & 0	&$\frac{1}{8}\pi $	&$\frac{2}{8}\pi $	&$\frac{3}{8}\pi $	&$\frac{4}{8}\pi $	&$\frac{5}{8}\pi $	&$\frac{6}{8}\pi $	&$\frac{7}{8}\pi $	&$\pi $\\
\hline
$P_{1,ideal}$	&	0.5	&	0.5	&	0.5	&	0.5	&	0.5	&	0.5	&	0.5	&	0.5	&	0.5	\\
$P_{2,ideal}$	&	0.5	&	0.5	&	0.5	&	0.5	&	0.5	&	0.5	&	0.5	&	0.5	&	0.5	\\
$P_{3,ideal}$	&	0	&	0	&	0	&	0	&	0	&	0	&	0	&	0	&	0	\\
$P_{4,ideal}$	&	0	&	0	&	0	&	0	&	0	&	0	&	0	&	0	&	0	\\
$P_{1,exp}$	&	0.4147	&	0.3918	&	0.3865	&	0.4289	&	0.4156	&	0.382	&	0.4058	&	0.4356	&	0.4123	\\
$P_{2,exp}$	&	0.4627	&	0.474	&	0.4679	&	0.4348	&	0.4359	&	0.4697	&	0.4507	&	0.4189	&	0.4416	\\
$P_{3,exp}$	&	0.0601	&	0.0671	&	0.0722	&	0.0782	&	0.0898	&	0.0877	&	0.0792	&	0.0933	&	0.0823	\\
$P_{4,exp}$	&	0.0625	&	0.0671	&	0.0734	&	0.0581	&	0.0587	&	0.0606	&	0.0642	&	0.0522	&	0.0639	\\
\hline
\end{tabular}

\begin{tabular}{lrrrrrrrrr}
\hline
Node/T     & 0	&$\frac{1}{8}\pi $	&$\frac{2}{8}\pi $	&$\frac{3}{8}\pi $	&$\frac{4}{8}\pi $	&$\frac{5}{8}\pi $	&$\frac{6}{8}\pi $	&$\frac{7}{8}\pi $	&$\pi $\\
\hline
$P_{1,ideal}$	&	0.5	&	0.125	&	0.25	&	0.625	&	0.5	&	0.125	&	0.25	&	0.625	&	0.5	\\
$P_{2,ideal}$	&	0.5	&	0.625	&	0.25	&	0.125	&	0.5	&	0.625	&	0.25	&	0.125	&	0.5	\\
$P_{3,ideal}$	&	0	&	0.125	&	0.25	&	0.125	&	0	&	0.125	&	0.25	&	0.125	&	0	\\
$P_{4,ideal}$	&	0	&	0.125	&	0.25	&	0.125	&	0	&	0.125	&	0.25	&	0.125	&	0	\\
$P_{1,exp}$	&	0.4178	&	0.1655	&	0.1969	&	0.4932	&	0.4729	&	0.1492	&	0.1977	&	0.4217	&	0.4332	\\
$P_{2,exp}$	&	0.3541	&	0.548	&	0.2749	&	0.1504	&	0.3824	&	0.6258	&	0.2503	&	0.1085	&	0.4308	\\
$P_{3,exp}$	&	0.1376	&	0.1015	&	0.211	&	0.1467	&	0.0594	&	0.1047	&	0.2316	&	0.2796	&	0.0734	\\
$P_{4,exp}$	&	0.0904	&	0.185	&	0.3171	&	0.2096	&	0.0853	&	0.1203	&	0.3205	&	0.1902	&	0.0626	\\
\hline
\end{tabular}

\begin{tabular}{lrrrrrrrrr}
\hline
Node/T     & 0	&$\frac{1}{8}\pi $	&$\frac{2}{8}\pi $	&$\frac{3}{8}\pi $	&$\frac{4}{8}\pi $	&$\frac{5}{8}\pi $	&$\frac{6}{8}\pi $	&$\frac{7}{8}\pi $	&$\pi $\\
\hline
$P_{1,ideal}$	&	0.25	&	0.5	&	0.25	&	0	&	0.25	&	0.5	&	0.25	&	0	&	0.25\\
$P_{2,ideal}$	&	0.25	&	0	&	0.25	&	0.5	&	0.25	&	0	&	0.25	&	0.5	&	0.25\\
$P_{3,ideal}$	&	0.25	&	0.5	&	0.25	&	0	&	0.25	&	0.5	&	0.25	&	0	&	0.25\\
$P_{4,ideal}$	&	0.25	&	0	&	0.25	&	0.5	&	0.25	&	0	&	0.25	&	0.5	&	0.25\\
$P_{1,exp}$&	0.2642	&	0.4163	&	0.2227	&	0.0172	&	0.2191	&	0.4136	&	0.2167	&	0.0591	&	0.2476\\
$P_{2,exp}$&	0.2724	&	0.0156	&	0.3	&	0.4678	&	0.2367	&	0.0227	&	0.2808	&	0.4864	&	0.199\\
$P_{3,exp}$&	0.2561	&	0.5525	&	0.2409	&	0.0043	&	0.3145	&	0.5545	&	0.2266	&	0.0227	&	0.3204\\
$P_{4,exp}$&	0.2073	&	0.0156	&	0.2364	&	0.5107	&	0.2297	&	0.0091	&	0.2759	&	0.4318	&	0.233\\
\hline
\end{tabular}

\begin{tabular}{lrrrrrrrrr}
\hline
Node/T     & 0	&$\frac{1}{8}\pi $	&$\frac{2}{8}\pi $	&$\frac{3}{8}\pi $	&$\frac{4}{8}\pi $	&$\frac{5}{8}\pi $	&$\frac{6}{8}\pi $	&$\frac{7}{8}\pi $	&$\pi $\\
\hline
$P_{1,ideal}$	&	0.25	&	0.625	&	0.5	&	0.125	&	0.25	&	0.625	&	0.5	&	0.125	&	0.25\\
$P_{2,ideal}$	&	0.25	&	0.125	&	0	&	0.125	&	0.25	&	0.125	&	0	&	0.125	&	0.25\\
$P_{3,ideal}$	&	0.25	&	0.125	&	0	&	0.125	&	0.25	&	0.125	&	0	&	0.125	&	0.25\\
$P_{4,ideal}$	&	0.25	&	0.125	&	0.5	&	0.625	&	0.25	&	0.125	&	0.5	&	0.625	&	0.25\\
$P_{1,exp}$&	0.2056	&	0.4226	&	0.3307	&	0.1129	&	0.168	&	0.4883	&	0.3425	&	0.1321	&	0.3347\\
$P_{2,exp}$&	0.1542	&	0.2469	&	0.0906	&	0.0887	&	0.127	&	0.1596	&	0.0827	&	0.0566	&	0.3431\\
$P_{3,exp}$&	0.2477	&	0.1674	&	0.0354	&	0.1411	&	0.3238	&	0.1643	&	0.0276	&	0.1358	&	0.1715\\
$P_{4,exp}$&	0.3925	&	0.1632	&	0.5433	&	0.6573	&	0.3811	&	0.1878	&	0.5472	&	0.6755	&	0.1506\\
\hline
\end{tabular}

\caption{The ideal theoretical and experimental probability distributions of CTQWs on $K_4$ graph with initial states
${\left| {{\varphi _{ini}}} \right\rangle _1} = \left[ {1,0,0,0} \right]'$,
${\left| {{\varphi _{ini}}} \right\rangle _2} = \frac{1}{{\sqrt 2 }}\left[ {1,1,0,0} \right]'$,
${\left| {{\varphi _{ini}}} \right\rangle _3} = \frac{1}{{\sqrt 2 }}\left[ {1, - 1,0,0} \right]'$,
${\left| {{\varphi _{ini}}} \right\rangle _4} = \frac{1}{{\sqrt 2 }}\left[ {1, - i,0,0} \right]'$,
${\left| {{\varphi _{ini}}} \right\rangle _5} = \frac{1}{ 2 }\left[ {1, i, 1, i} \right]'$,
${\left| {{\varphi _{ini}}} \right\rangle _6} = \frac{1}{ 2 }\left[ {1, i, i, -1} \right]'$
shown in six sub-tables  from top to bottom. For each sub-table, the first four rows are ideal results and the last four rows are experimental results.}

\end{table*}


\begin{thebibliography}{10}

\bibitem{farhi1998quantum}
Farhi, E. \& Gutmann, S.
\newblock Quantum computation and decision trees.
\newblock {\em Phys. Rev. A} \textbf{58,} 915 (1998).

\bibitem{kempe2003quantum}
Kempe, J.
\newblock Quantum random walks: an introductory overview.
\newblock {\em Contemp. Phys.} \textbf{44,} 307-327 (2003).

\bibitem{childs2013universal}
Childs, A. M., Gosset, D. \& Webb, Z.
\newblock Universal computation by multiparticle quantum walk.
\newblock {\em Science} \textbf{339,} 791-794 (2013).

\bibitem{childs2004spatial}
Childs, A. M. \& Goldstone, J.
\newblock Spatial search by quantum walk.
\newblock {\em Phys. Rev. A} \textbf{70,} 022314 (2004).

\bibitem{douglas2008}
Douglas, B. L. \&  Wang, J. B. 
\newblock A classical approach to the graph isomorphism problem using quantum walks.
\newblock {\em J. Phys. A} \textbf{41,} 075303 (2008).


\bibitem{gamble2010two}
Gamble, J. K., Friesen, M., Zhou, D., Joynt, R. \& Coppersmith, S. N.
\newblock Two-particle quantum walks applied to the graph isomorphism problem.
\newblock {\em Phys. Rev. A} \textbf{81,} 052313 (2010).


\bibitem{berry2011}
Berry, S. D. \&  Wang, J. B. 
\newblock Two-particle quantum walks: Entanglement and graph isomorphism testing.
\newblock {\em Phys. Rev. A} \textbf{83,} 042317 (2011).


\bibitem{Berry2010}
Berry, S. D. \&  Wang, J. B. 
\newblock Quantum-walk-based search and centrality.
\newblock {\em Phys. Rev. A} \textbf{82,} 042333 (2010).

\bibitem{sanchez2012quantum}
S{\'a}nchez-Burillo, E., Duch, J., G{\'o}mez-Garde{\~n}es, J. \& Zueco, D.
\newblock Quantum navigation and ranking in complex networks.
\newblock {\em Sci. Rep.} \textbf{2,} 605 (2012).

\bibitem{lloyd1996universal}
Lloyd, S.
\newblock Universal quantum simulators.
\newblock {\em Science} \textbf{273,} 1073-1078 (1996).

\bibitem{berry2012black}
Berry, D. W. \& Childs, A. M.
\newblock Black-box hamiltonian simulation and unitary implementation.
\newblock {\em Quantum Inf. Comput.} \textbf{12,} 29-62 (2012).


\bibitem{schreiber20122d}
Schreiber, A. {\em et al.}
\newblock A 2D quantum walk simulation of two-particle dynamics.
\newblock {\em Science} \textbf{336,} 55-58 (2012).

\bibitem{engel2007evidence}
Engel, G. S. {\em et al.}
\newblock Evidence for wavelike energy transfer through quantum coherence in photosynthetic systems.
\newblock {\em Nature} \textbf{446,} 782-786 (2007).

\bibitem{rebentrost2009environment}
Rebentrost, P. {\em et al.}
\newblock Environment-assisted quantum transport.
\newblock {\em New J. Phys.} \textbf{11,} 033003 (2009).


\bibitem{du2003experimental}
Du, J. {\em et al.}
\newblock Experimental implementation of the quantum random-walk algorithm.
\newblock {\em Phys. Rev. A} \textbf{67,} 042316 (2003).

\bibitem{ryan2005experimental}
Ryan, C. A. {\em et al.}
\newblock Experimental implementation of a discrete-time quantum random walk on an NMR quantum-information processor.
\newblock {\em Phys. Rev. A} \textbf{72,} 062317 (2005).

\bibitem{do2005experimental}
Do, B. {\em et al.}
\newblock Experimental realization of a quantum quincunx by use of linear
  optical elements.
\newblock {\em J. Opt. Soc. Am. B} \textbf{22,} 499-504 (2005).

\bibitem{schreiber2010photons}
Schreiber, A. {\em et al.}
\newblock Photons walking the line: a quantum walk with adjustable coin operations.
\newblock {\em Phys. Rev. Lett.} \textbf{104,} 050502 (2010).

\bibitem{xue2009quantum}
Xue, P., Sanders, B. C. \& Leibfried, D.
\newblock Quantum walk on a line for a trapped ion.
\newblock {\em Phys. Rev. Lett.} \textbf{103,} 183602 (2009).

\bibitem{schmitz2009quantum}
Schmitz, H. {\em et al.}
\newblock Quantum walk of a trapped ion in phase space.
\newblock {\em Phys. Rev. Lett.} \textbf{103,} 090504 (2009).

\bibitem{zahringer2010realization}
Z{\"a}hringer, F. {\em et al.}
\newblock Realization of a quantum walk with one and two trapped ions.
\newblock {\em Phys. Rev. Lett.} \textbf{104,} 100503 (2010).

\bibitem{karski2009quantum}
Karski, M. {\em et al.}
\newblock Quantum walk in position space with single optically trapped atoms.
\newblock {\em Science} \textbf{325,} 174-177 (2009).


\bibitem{perets2008realization}
Perets, H. B. {\em et al.}
\newblock Realization of quantum walks with negligible decoherence in waveguide lattices.
\newblock {\em Phys. Rev. Lett.} \textbf{100,} 170506 (2008).


\bibitem{carolan2014experimental}
Carolan, J. {\em et al.}
\newblock On the experimental verification of quantum complexity in linear optics.
\newblock {\em Nat. Photonics} \textbf{8,} 621 (2014).

\bibitem{qwbook2014}
Manouchehri, K. \& Wang, J. B. 
\newblock {\em Physical Implementation of Quantum Walks}.
\newblock Springer-Verlag, Berlin, (2014).

\bibitem{qubitImplementationQW}
Exceptions, such as~\cite{du2003experimental}, adopted the qubit model, but there was no discussion on potentially efficient implementation of quantum walks.

\bibitem{ng2004iterative}
Ng, M. K.
\newblock {\em Iterative Methods for Toeplitz Systems}.
\newblock Oxford University Press New York, (2004).

\bibitem{aharonov2003adiabatic}
Aharonov, D. \& Ta-Shma, A.
\newblock Adiabatic quantum state generation and statistical zero knowledge.
\newblock In {\em Proceedings of the thirty-fifth annual ACM symposium on
  Theory of computing} 20-29. ACM (2003).
  
\bibitem{berry2007efficient}
Berry, D. W., Ahokas, G., Cleve, R. \& Sanders, B. C.
\newblock Efficient quantum algorithms for simulating sparse hamiltonians.
\newblock {\em Commun. Math. Phys.} \textbf{270,} 359-371 (2007).

\bibitem{childs2009limitations}
Childs, A. M. \& Kothari, R.
\newblock Limitations on the simulation of non-sparse hamiltonians. 
\newblock {\em Quantum Inf. Comput.} \textbf{10,} 669-684 (2009).

\bibitem{childs2010relationship}
Childs, A. M.
\newblock On the relationship between continuous-and discrete-time quantum
  walk.
\newblock {\em Commun. Math. Phys.} \textbf{294,} 581-603 (2010).

\bibitem{gray2006toeplitz}
Gray, R. M.
\newblock {\em Toeplitz and Circulant Matrices: A Review}.
\newblock Now Publishers Inc., (2006).

\bibitem{buhrman2001quantum}
Buhrman, H., Cleve, R., Watrous, J. \& De~Wolf, R.
\newblock Quantum fingerprinting.
\newblock {\em Phys. Rev. Lett.} \textbf{87,} 167902 (2001).

\bibitem{han2010folding}
Han, D. {\em et al.}
\newblock Folding and cutting DNA into reconfigurable topological nanostructures.
\newblock {\em Nat. Nanotechnol} \textbf{5,} 712-717 (2010).


\bibitem{peres1984stability}
Peres, A.
\newblock Stability of quantum motion in chaotic and regular systems.
\newblock {\em Phys. Rev. A} \textbf{30,} 1610 (1984).

\bibitem{prosen2002stability}
Prosen, T. \& Znidaric, M.
\newblock Stability of quantum motion and correlation decay.
\newblock {\em J. Phys. A: Math. Gen.} \textbf{35,} 1455 (2002).

\bibitem{aaronson2011computational}
Aaronson, S. \& Arkhipov, A.
\newblock The computational complexity of linear optics.
\newblock In {\em Proceedings of the forty-third annual ACM symposium on Theory
  of computing} 333--342. ACM (2011).
  
\bibitem{bremner2010classical}
Bremner, M. J., Jozsa, R. \& Shepherd, D. J. 
\newblock Classical simulation of commuting quantum computations implies
  collapse of the polynomial hierarchy.
\newblock {\em Proceedings of the Royal Society A: Mathematical, Physical and
  Engineering Science}, rspa.2010.0301 (2010).

\bibitem{bremner2015average}
Bremner, M. J., Montanaro, A. \& Shepherd, D. J.
\newblock Average-case complexity versus approximate simulation of commuting
  quantum computations, Preprint at
\newblock \url{http://arxiv.org/abs/1504.07999} (2015).

\bibitem{nielsen2010quantum}
Nielsen, N. A. \& Chuang, I. L.
\newblock {\em Quantum Computation and Quantum Information}.
\newblock Cambridge University Press, (2010).

\bibitem{childs2004quantum}
Childs, A. M.
\newblock {\em Quantum information processing in continuous time}.
\newblock PhD Thesis, Massachusetts Institute of Technology (2004).

\bibitem{shepherd2009temporally}
Shepherd, D. \& Bremner, M. J.
\newblock Temporally unstructured quantum computation.
\newblock {\em Proceedings of the Royal Society A: Mathematical, Physical and
  Engineering Science} \textbf{465,} 1413-1439 (2009).
  
\bibitem{fujii13}
Fujii, K. \& Morimae, T.
\newblock Quantum commuting circuits and complexity of {I}sing partition
  functions, Preprint at
\newblock \url{http://arxiv.org/abs/1311.2128} (2013).

\bibitem{goldberg14}
Goldberg, L. A. \& Guo, H.
\newblock The complexity of approximating complex-valued {Ising} and {Tutte}
  partition functions, Preprint at
\newblock \url{http://arxiv.org/abs/1409.5627} (2014).

\bibitem{stockmeyer1985on}
Stockmeyer, L. J.
\newblock On approximation algorithms for {\#}{P}.
\newblock {\em SIAM J. Comput.} \textbf{14,} 849-861 (1985).

\bibitem{papadimitriou1994computational}
Papadimitriou, C.
\newblock {\em Computational {C}omplexity}.
\newblock Addison-Wesley, (1994).

\bibitem{van2011simulating}
Nest, M.
\newblock Simulating quantum computers with probabilistic methods.
\newblock {\em Quantum Inf. Comput.} \textbf{11,} 784-812 (2011).

\bibitem{schwarz2013simulating}
Schwarz, M. \& Nest, M.
\newblock Simulating quantum circuits with sparse output distributions, Preprint at
\newblock \url{http://arxiv.org/abs/1310.6749} (2013).

\bibitem{zhou2011adding}
Zhou, X. Q. {\em et al.} 
\newblock Adding control to arbitrary unknown quantum operations.
\newblock {\em Nat. Commun.} \textbf{2,} 413 (2011).

\bibitem{knill1998power}
Knill, E., \& Laflamme, R.
\newblock Power of one bit of quantum information.
\newblock {\em Phys. Rev. Lett.} \textbf{81,} 5672 (1998).

\bibitem{rohde2012error}
Rohde, P. P. \& Ralph, T. C.
\newblock Error tolerance of the boson-sampling model for linear optics quantum
  computing.
\newblock {\em Phys. Rev. A} \textbf{85,} 022332 (2012).

\bibitem{rangarajan2009optimizing}
Rangarajan, R., Goggin, M. \& Kwiat, P.
\newblock Optimizing type-I polarization-entangled photons.
\newblock {\em Opt. Express} \textbf{17,} 18920-18933 (2009). 

\end{thebibliography}

\begin{thebibliography}{1}

\bibitem{ng2004iterative_appendix}
Ng, M. K.
\newblock {\em Iterative Methods for Toeplitz Systems}.
\newblock Oxford University Press New York, (2004).

\bibitem{zhou2011self}
Zhou, H.
\newblock On self-complementary of circulant graphs.
\newblock In {\em Theoretical and Mathematical Foundations of Computer
  Science} 464-471, Springer (2011).

\bibitem{rajasingh2013spanning}
Rajasingh, I. \& Natarajan, P.
\newblock Spanning trees in circulant networks.
\newblock 2013.

\bibitem{mosca2004exact}
Mosca, M. \& Zalka, C.
\newblock Exact quantum fourier transforms and discrete logarithm algorithms.
\newblock {\em Int. J. Quantum Inf.} \textbf{2} 91-100 (2004).

\bibitem{hales2000improved}
Hales, L. \& Hallgren, S.
\newblock An improved quantum fourier transform algorithm and applications.
\newblock In {\em Foundations of Computer Science, 2000. Proceedings. 41st
  Annual Symposium on} 515-525, IEEE (2000).

\bibitem{childs2004quantum}
Childs, A. M.
\newblock {\em Quantum information processing in continuous time}.
\newblock PhD Thesis, Massachusetts Institute of Technology (2004).

\end{thebibliography}
\end{document}